# Thermographic measurements of spin-current-induced temperature modulation in metallic bilayers

R. Iguchi[1,*], A. Yagmur[1,2,*], Y.-C. Lau[1,3], S. Daimon[2,4,5], E. Saitoh[2,4-7], M. Hayashi[1,3], and K. Uchida[1,6,8,†]

[1]*National Institute for Materials Science, Tsukuba 305-0047, Japan*

[2]*Institute for Materials Research, Tohoku University, Sendai 980-8577, Japan*

[3]*Department of Physics, The University of Tokyo, Tokyo 113-0033, Japan*

[4]*Department of Applied Physics, The University of Tokyo, Tokyo 113-8656, Japan*

[5]*Advanced Institute for Materials Research, Tohoku University, Sendai 980-8577, Japan*

[6]*Center for Spintronics Research Network, Tohoku University, Sendai 980-8577, Japan*

[7]*Advanced Science Research Center, Japan Atomic Energy Agency, Tokai 319-1195, Japan*

[8]*Department of Mechanical Engineering, The University of Tokyo, Tokyo 113-8656, Japan*

Spin-to-heat current conversion effects have been investigated in bilayer films consisting of a paramagnetic metal (PM; Pt, W, or Ta) and a ferromagnetic metal (FM; CoFeB or permalloy). When a charge current is applied to the PM/FM bilayer film, a spin current is generated across the PM/FM interface owing to the spin Hall effect in PM. The spin current was found to exhibit cooling and heating features depending on the sign of the spin Hall angle of PM, where the spin-current-induced contribution is estimated by subtracting the contribution of the anomalous Ettingshausen effect in FM monolayer films. We also found that the magnitude of the spin-current-induced temperature modulation in the Pt/CoFeB film is greater than but comparable to that in the Pt/permalloy film, although the spin dependence of the Peltier coefficient for CoFeB is expected to be greater than that for permalloy. We discuss the origin of the observed behaviors with the aid of model calculations; the signals in the PM/FM films may contain the contributions not only from the electron-driven spin-dependent Peltier effect but also from the magnon-driven spin Peltier effect.

[*] These authors contributed equally to this work.

[†] UCHIDA.Kenichi@nims.go.jp

## I. INTRODUCTION

The field of spin caloritronics aims to develop novel physics and applications based on the interplay between spintronics and thermal transport effects [1–6]. Experimental studies on spin caloritronics begin with the investigation of heat-to-spin current conversion phenomena. One of such phenomena is the spin Seebeck effect (SSE), which refers to the generation of a spin current as a result of a heat current in magnetic materials [7–20]. Since the SSE appears in magnetic insulators, this phenomenon is now understood in terms of non-equilibrium thermal magnon transport, and most of the experimental behaviors are explained by the magnon-based models [21–30]. In addition to the magnon-driven SSE, the heat-to-spin current conversion can arise also from conduction-electrons' spin transport; this is called the spin-dependent Seebeck effect (SdSE) because it originates from the difference in Seebeck coefficients between up- and down-spin electrons. After the pioneering work by Slachter *et al.*, the SdSE has been investigated in several ferromagnetic metals [31–34].

Since 2012, the spin caloritronics research has entered the investigation of the inverse effects: the spin-to-heat current conversion phenomena. This stream is triggered by the observation of the spin-dependent Peltier effect (SdPE), the Onsager reciprocal of the SdSE, in ferromagnetic metal (FM)/paramagnetic metal (PM)/FM pillar structures by Flipse *et al.* [35]. In 2014, they also reported the observation of the spin Peltier effect (SPE) in Pt/ferrimagnetic insulator [yttrium iron garnet (YIG)] junctions by using micro-fabricated thermopile sensors [36]. The SdPE and SPE refer to the generation of a heat current as a result of spin-current injection and, in analogy with the heat-to-spin current conversion phenomena, the mechanism of the SdPE (SPE) is discussed in terms of non-equilibrium transport of conduction-electrons' spins (magnons) [Figs. 1(a) and 1(b)]. However, the experimental research on the spin-to-heat current conversion phenomena is limited to a few studies [37–41], and their behaviors and mechanisms are not sufficiently investigated. This situation is attributed mainly to difficulty in measuring the SdPE and SPE; the spin-current-induced temperature change appears in nanoscale thin film devices and its magnitude is typically smaller than 10 mK [35]. The conventional temperature measurements in such nanoscale devices also have difficulty in quantitative estimation of the spin-to-heat current conversion efficiency because the temperature modulation concomitant with spin currents is confined near heat-source positions [38].

To overcome this situation, we have recently established a versatile method for measuring the SPE based on the lock-in thermography (LIT) technique [38–40]. This method allows imaging of the temperature modulation induced by the SPE with high temperature and spatial resolutions (< 0.1 mK and < 10 $\mu$m) and requires no micro-fabrication processes, realizing systematic investigations of the spin-to-heat current conversion properties. In Refs. [38,40], by using the LIT method, we have systematically investigated the temperature modulation induced by the SPE in PM/YIG junctions and revealed its unconventional spatial distribution. However, the investigation of the spin-to-heat current conversion phenomena using the LIT method has been performed only for magnetic insulators, where the spin-to-heat current conversion arises only from the magnon-driven SPE because of the absence of the conduction-electrons' contribution.

In this work, we have applied the LIT method to PM/FM bilayer films and investigated the spin-to-heat current conversion phenomena in metallic systems. The spin-to-heat current conversion in metallic systems is more complicated than that in insulating systems since it can be driven by both the electron-driven SdPE and magnon-driven SPE and be contaminated by thermoelectric effects in FM. The systematic measurements based on the LIT provide a crucial piece of information for separating these contributions and clarifying the spin-to-heat current conversion mechanisms in metals.

This paper is organized as follows. In Sec. II, we explain the details of the experimental procedures and configurations for the measurements of the spin-to-heat current conversion phenomena using the LIT method. In Sec. III, we report the observation of the spin-current-induced temperature modulation in PM/FM bilayer films, followed by model calculations to discuss the origin of the observed behaviors. The last Sec. IV is devoted to the conclusion of the present study.

## II. EXPERIMENTAL PROCEDURE AND CONFIGURATION

The sample system used in this study consists of a PM film formed on a FM film. Here, we select two different FM materials. The first one is $Co_{20}Fe_{60}B_{20}$ (CoFeB), which is known to have large difference in spin-dependent Seebeck/Peltier coefficients [42,43]. The other one is $Ni_{81}Fe_{19}$ [permalloy (Py)], which is a typical FM with moderate difference in spin-dependent Seebeck/Peltier coefficients [44]. As the PM layer, we select Pt, W, and Ta since they have strong spin-orbit coupling, of which the sign for Pt is opposite to that for W and Ta. The thickness of the PM (FM) layer is 10 nm (20 nm) except for the samples used for the measurements of the thickness dependence shown in Sec. IIIB. The PM/FM bilayer films were fabricated on sapphire substrates and patterned into U-shaped structure by sputtering the PM and FM layers through a metallic shadow mask [Figs. 2(a) and 2(b)], where the line-width of the U-shaped structure is 0.2 mm and the total line length of U-shaped structure is $l_{tot}$ = 4.6 mm. To avoid the oxidation, Ta(1 nm)/MgO(2 nm) protective layers were sputtered on the PM layer.

In the PM/FM bilayer film, both the electron-driven SdPE and magnon-driven SPE can contribute to the spin-to-heat current conversion. To excite the SdPE and SPE in the PM/FM system, we employ the spin Hall effect (SHE) [45–49] in the PM layer for injecting a spin current into the FM layer [Figs. 1(a) and 1(b)]. When a charge current $\mathbf{J}_c$ with its density vector $\mathbf{j}_c$ flows in the PM layer of the PM/FM system along the $y$ direction, a spin current $\mathbf{J}_s$ with its density vector $\mathbf{j}_s$ and the spin-polarization vector $\boldsymbol{\sigma}$ is generated due to the SHE in PM and injected into FM. Here, electrons with $\boldsymbol{\sigma}$ along the $x$ direction induce $\mathbf{J}_s$ along the $z$ direction, since the SHE holds the following relation

$$\mathbf{j}_s = \theta_{SH} \mathbf{j}_c \times \boldsymbol{\sigma}, \tag{1}$$

where $\theta_{SH}$ is the spin Hall angle of PM. In the SdPE (SPE), the spin current in FM is carried by conduction electrons (magnons). When the $\boldsymbol{\sigma}$ direction is parallel or antiparallel to the magnetization $\mathbf{M}$ of FM, the spin current induces a temperature gradient along the stacking direction, i.e., the $z$ direction. Here, the magnitude of the temperature gradient is proportional to $|\mathbf{J}_s|$ and its direction is dependent on the $\boldsymbol{\sigma}$ direction and the sign of the SdPE or SPE coefficient. As shown in Figs. 1(a) and 1(b), the symmetry of the SHE-driven SdPE is the same as that of the SPE; both effects can be superimposed. To realize the detection of the SdPE and SPE in the PM/FM system, it is also important to distinguish their signals from the anomalous Ettingshausen effect (AEE), which is a transverse thermoelectric effect occurring in FM [50]. Since the temperature gradient due to the AEE in FM is generated in the direction of the cross product of $\mathbf{J}_c$ and $\mathbf{M}$, it contaminates the SdPE and SPE signals in the PM/FM bilayer systems [Fig. 1(c)]. We separate the spin-current-induced signals from the AEE signals by comparing the results in the PM/FM systems with those in FM monolayer films, where only the AEE contribution exists (see Sec. IIIA for details).

To detect the temperature change induced by the spin current in the PM/FM samples, we performed the LIT measurements at room temperature and atmospheric pressure [51,52]. First of all, the surface of the samples was coated with insulating black ink to enhance infrared emissivity. In the LIT measurements, we measured the spatial distribution of infrared radiation thermally emitted from the surface of the U-shaped PM/FM films with applying a rectangularly-modulated AC voltage with the amplitude $V$, frequency $f$, and zero DC offset to the films [Fig. 2(a)]. In this study, we fixed the lock-in frequency at $f$ = 5 Hz. By extracting the first harmonic response of detected thermal images via Fourier analyses, we can obtain the lock-in amplitude $A$ and phase $\phi$ images, enabling highly-sensitive detection of thermo-spin and thermoelectric effects free from the Joule-heating background [Fig. 2(a)] [38,40]. Here, the $A$ ($\phi$) image provides the spatial distribution of the magnitude of the voltage-induced temperature modulation (the sign of the temperature modulation as well as the time delay due to thermal diffusion), where the $A$ ($\phi$) values are defined in the ranges of $A \geq 0$ ($0° \leq \phi < 360°$). During the LIT measurements, to saturate the magnetization **M** of the CoFeB and Py films along the magnetic field **H**, we applied an in-plane magnetic field $H$ with the magnitude of $|H| > 0.5$ kOe along the $x$ direction [see the magnetization curve of the CoFeB film shown in Fig. 2(c)]. To extract the pure SdPE, SPE, and AEE contributions, which reverse sign by reversing **H**, we calculated the $A_{\text{odd}}$ and $\phi_{\text{odd}}$ images showing the distribution of the voltage-induced temperature modulation with the $H$-odd dependence. Here, the $A_{\text{odd}}$ and $\phi_{\text{odd}}$ images are obtained by subtracting the LIT images at $H < -0.5$ kOe from those at $H > +0.5$ kOe and dividing the subtracted images by 2. In our samples, owing to the U-shaped structure, the symmetries of the SdPE, SPE, and AEE can be confirmed simultaneously because the relative orientation of $\mathbf{J}_c$ and **M** is different between the areas L, R, and C, where $\mathbf{J}_c \perp \mathbf{M}$ on L and R and $\mathbf{J}_c \parallel \mathbf{M}$ on C when **M** is along the $x$ direction [Fig. 2(b)]. Therefore, the temperature modulation due to the SdPE, SPE, and AEE appears on L and R, while it disappears on C [38,40]. Since the $\mathbf{J}_c$ direction on L is opposite to that on R, the sign of the temperature modulation induced by these phenomena is reversed between these areas.

## III. RESULTS AND DISCUSSION

### A. Separation of spin-current-induced temperature modulation from anomalous Ettingshausen effect in PM/CoFeB systems

Figures 2(d) and 2(e) respectively show the $A_{\text{odd}}$ and $\phi_{\text{odd}}$ images for the Pt/CoFeB film at $V = 10$ V and $|H| = 1.4$ kOe, where $V = 10$ V corresponds to the electric field magnitude $E$ of 2.2 kV/m and the charge-current amplitude of 20 mA for this sample. We observed clear temperature-modulation signals on L and R, where $\mathbf{J}_c \perp \mathbf{M}$, and ~180° difference in $\phi$ between L and R, while the signals disappear on C, where $\mathbf{J}_c \parallel \mathbf{M}$. Since the heat-conduction condition is the same for L and R, this $\phi$ shift is irrelevant to the time delay caused by thermal diffusion, indicating that the sign of the temperature modulation is reversed depending on the direction of $\mathbf{J}_c$. In Figs. 2(f) and 2(g), we show the $V$ dependence of $A_{\text{odd}}$ and $\phi_{\text{odd}}$ in the Pt/CoFeB film, respectively. The $A_{\text{odd}}$ value is proportional to $V$, while the $\phi_{\text{odd}}$ shift of ~180° remains unchanged with respect to $V$. These behaviors are in good agreement with the features of the SPE, SdPE, and AEE [35,36,38–40,50].

To clarify the origin of the temperature modulation in the Pt/CoFeB film, we performed the control experiments using a CoFeB monolayer film, without the PM layer, and a W/CoFeB (Ta/CoFeB) bilayer film in which the Pt layer is replaced with the W (Ta) layer. As shown in Figs. 3(a) and 3(b), we found that the CoFeB, W/CoFeB, and Ta/CoFeB films exhibit the clear temperature modulation with the same symmetry and sign as those for the Pt/CoFeB film. In contrast, the signal magnitude depends on the sample species; the $A_{odd}$ values on L and R for the Pt/CoFeB film (W/CoFeB and Ta/CoFeB films) are greater (smaller) than those for the CoFeB monolayer, which contain only the AEE contribution [Figs. 3(a) and 3(c)]. This result indicates that the positive (negative) spin-current-induced contribution driven by the SHE in Pt (W and Ta) is superimposed on the positive AEE background in the CoFeB layer, since the sign of $\theta_{SH}$ in Pt (W and Ta) is positive (negative) and the PM layer exhibits no AEE (note that the $H$-linear contribution of the ordinary Ettingshausen effect in PM is negligibly small [40,50]). Importantly, during the LIT measurements, we fixed the amplitude of the voltage $V$, not the charge current, applied to the PM/CoFeB and CoFeB films; if we regard the PM/CoFeB bilayer film as a simple parallel circuit comprising the PM and CoFeB layers with negligible interface resistivity [53], the charge-current density and resultant AEE contribution in the CoFeB layer of the PM/CoFeB bilayers is the same as that in the CoFeB monolayer. Based on this interpretation, we estimate the spin-current-induced contribution in the PM/CoFeB films by subtracting the signal in the CoFeB monolayer from that in the PM/CoFeB bilayers. As shown in Fig. 3(d), the subtracted LIT amplitude per unit electric field $\Delta A_{odd}/E$ with $E = V/l_{tot}$ in the Pt/CoFeB film (W/CoFeB and Ta/CoFeB films) exhibit the clear positive (negative) contribution, consistent with the characteristic of the SPE and SdPE. Here, the sign of the spin-current-induced signal in the Pt/CoFeB film is the same as that of the SPE signal in the Pt/YIG system [38].

**B. Thickness dependence**

In this subsection, we show the thickness dependence of the voltage-induced temperature modulation. First, to further support our interpretation that the $\Delta A_{odd}$ signals in the PM/CoFeB films originate from the SHE in the PM layer, we investigated the PM-layer thickness dependence of the temperature modulation. Here, we used the W/CoFeB films with the different W-layer thickness $d_W$ and the constant CoFeB-layer thickness of 20 nm. Figures 4(a) and 4(b) show the $A_{odd}$ and $\phi_{odd}$ images for the CoFeB monolayer and W($d_W$)/CoFeB films at $V = 10$ V and $|H| = 0.7$ kOe. We observed clear temperature-modulation signals with the aforementioned features in all the films. As shown in Fig. 4(c), the magnitude of the $A_{odd}$ signals in the W($d_W$)/CoFeB films is smaller than that in the CoFeB monolayer film. To quantitatively estimate the $d_W$ dependence of the signal reduction, $\Delta A_{odd}$, we normalized the $\Delta A_{odd}$ signals by the charge-current density $j_c^W$ in the W layer, based on the parallel circuit model [53]. As shown in Fig. 4(d), the magnitude of $\Delta A_{odd}/j_c^W$ for the W(5 nm)/CoFeB film is much greater than that for the W(10 or 15 nm)/CoFeB films and the resistivity of the 5-nm-thick W film is much greater than that of the 10- and 15-nm-thick films. This behavior is consistent with the W-thickness dependence of $\theta_{SH}$; the SHE in W is known to be enhanced with decreasing the thickness due to the contribution from the highly-resistive $\beta$-W phase [54–56]. The W-thickness dependence observed here buttresses our basis that the difference in the temperature modulation between the PM/CoFeB bilayer and CoFeB monolayer films is attributed to the spin-current injection induced by the SHE.

Next, we measured the FM-layer thickness dependence of the spin-current-induced temperature modulation to investigate the length scale of the observed phenomena. To do this, we performed the same experiments using the Pt/CoFeB and CoFeB films with varying the CoFeB thickness $d_{CoFeB}$ while fixing the Pt thickness at 10 nm. Figure 5(a) shows the $d_{CoFeB}$ dependence of $A_{odd}/E$ for the Pt/CoFeB($d_{CoFeB}$) and CoFeB($d_{CoFeB}$) films at $|H| = 1.4$ kOe. We observed clear temperature-modulation signals in all the films and found that the magnitude of $A_{odd}/E$ monotonically increases with increasing $d_{CoFeB}$. The AEE signal in the CoFeB monolayer films exhibits an almost linear dependence on $d_{CoFeB}$; this behavior can be explained simply by the facts that the out-of-plane heat current induced by the AEE is constant in the CoFeB layer and that the resultant temperature difference is proportional to the integral of the heat current over the CoFeB thickness. In contrast, the $d_{CoFeB}$ dependence of the spin-current-induced signal in the Pt/CoFeB films, extracted by subtracting the AEE contributions in the CoFeB layer, shows a different behavior; as shown in Fig. 5(b), the magnitude of $\Delta A_{odd}/j_c^{Pt}$ in the Pt/CoFeB films gradually increases with increasing $d_{CoFeB}$ but saturates when $d_{CoFeB} > 30$ nm, where $j_c^{Pt}$ denotes the charge-current density in the Pt layer. This saturation behavior is qualitatively similar to the ferromagnetic- or ferrimagnetic-layer thickness dependence of the thermo-spin effects, such as the SSE, SdSE, and SPE [25,40,57]. In Sec. IIID, we discuss the origin of the $d_{CoFeB}$ dependence of the spin-current-induced signals by using model calculations.

## C. Comparison between Pt/CoFeB and Pt/Py systems

The above experiments clearly show that the PM/CoFeB films exhibit the spin-current-induced temperature modulation. However, the temperature modulation may include both the electron-driven SdPE and magnon-driven SPE contributions in the metallic samples. To obtain a clue for distinguishing the SdPE and SPE contributions, we measured the spin-current-induced temperature modulation also in the Pt/Py film under the same conditions as the CoFeB experiments. Since the SdPE coefficient of Py is believed to be much smaller than that of CoFeB [43,44,58], the SdPE contribution in the Pt/Py films is expected to be smaller than that in the Pt/CoFeB films.

In Figs. 6(a) and 6(b), we show the $A_{\text{odd}}$ and $\phi_{\text{odd}}$ images for the Py monolayer and Pt/Py bilayer films at $V =$ 10 V and $|H| = 1.4$ kOe. Both the samples exhibit clear temperature-modulation signals on L and R in the same manner as the CoFeB experiments, where the sign of the signals is reversed between L and R [Fig. 6(b)] and the magnitude is proportional to $V$ [Fig. 6(c)]. Importantly, the signal magnitude in the Pt/Py bilayer film was found to be greater than that in the Py monolayer film, indicating the finite spin-current contribution in the Pt/Py film. As shown in the inset to Fig. 6(c), the $\Delta A_{\text{odd}}$ signal in the Pt/Py film is proportional to $V$, consistent with the characteristic of the SPE and SdPE. The sign of the spin-current-induced temperature modulation in the Pt/Py film is the same as that in the Pt/CoFeB film.

Here, we compare the magnitude of the spin-current-induced temperature modulation between the Pt/CoFeB and Pt/Py films. The values of $\Delta A_{\text{odd}}/j_c^{\text{Pt}}$ on L for the Pt(10nm)/CoFeB(20nm) and Pt(10nm)/Py(20nm) films are estimated to be $0.71 \times 10^{-13}$ Km$^2$A$^{-1}$ and $0.32 \times 10^{-13}$ Km$^2$A$^{-1}$, respectively. The magnitude of the spin-current-induced signal in the Pt/CoFeB film is greater than but comparable to that in the Pt/Py film despite the substantial difference in electron-transport properties between CoFeB and Py [43,44]. Furthermore, as discussed in Sec. IIID, the magnitude of the spin-current-induced temperature modulation observed here is too large to be explained only by the SdPE contribution. These facts imply that not only the electron-driven SdPE but also the magnon-driven SPE contributes to the temperature modulation in our PM/FM bilayer films.

### D. Modeling of SdPE- and SPE-induced temperature modulation

To further discuss the origin of the observed spin-current-induced temperature modulation, we model the SHE-induced SdPE and SPE in the PM/FM bilayer films. The spin currents in FM are composed of conduction electrons and magnons. For conduction electrons in FM, the diffusive spin current is driven by the gradient of the spin-dependent electrochemical potentials $\mu_\sigma$ with the spin index $\sigma$ $(= \uparrow, \downarrow)$ as follows:

$$j_s = -\left(\sigma_\uparrow \nabla \frac{\mu_\uparrow}{e} - \sigma_\downarrow \nabla \frac{\mu_\downarrow}{e}\right) = -\frac{\sigma_{\text{FM}}}{2} \nabla \frac{\mu_s}{e} - \sigma_{\text{FM}} P_{\text{FM}} \nabla \frac{\mu_c}{e}, \tag{2}$$

where $\mu_s = \mu_\uparrow - \mu_\downarrow$, $\mu_c = (\mu_\uparrow + \mu_\downarrow)/2$, $e$ is the elemental charge, $\sigma_{\text{FM}} = \sigma_\uparrow + \sigma_\downarrow$ the electrical conductivity, and $P_{\text{FM}}$ the spin polarization of conduction electrons: $P_{\text{FM}} = (\sigma_\uparrow - \sigma_\downarrow)/\sigma_{\text{FM}}$. When no charge current exists along the spin current, this spin current gives rise to the SdPE-induced temperature modulation: $\Delta T_{\text{SdPE}} \propto -\Pi_s j_s$ with the SdPE coefficient $\Pi_s$, which is determined by the difference in the Peltier coefficient between the up- and down-spin conduction electrons: $\Pi_s = \Pi_\uparrow - \Pi_\downarrow$. Magnons can also be driven by the gradient of its accumulation $\mu_m$, and the magnon current is given by

$$j_m = -\sigma_m \nabla \frac{\mu_m}{e}, \tag{3}$$

where $\sigma_m$ is the magnon conductivity. The SPE-induced temperature change is described as $\Delta T_{SPE} \propto \Pi_{SPE} j_m$ with the SPE coefficient $\Pi_{SPE}$.

To estimate the SdPE- and SPE-induced temperature modulations, we determined the spin-current density by solving the diffusion equations for $\mu_s$, $\mu_c$, and $\mu_m$:

$$\nabla^2 \mu_s = \mu_s/\lambda^2$$
$$\nabla^2 \mu_c = 0 \tag{4}$$
$$\nabla^2 \mu_m = \mu_m/\lambda_m^2,$$

where $\lambda$ ($\lambda_m$) is the spin (magnon) diffusion length. We consider one-dimensional spin and magnon transports in the direction perpendicular to the PM/FM interface (the $z$ direction). The FM (PM) layer possesses the conductivity $\sigma_{FM(PM)}$, spin diffusion length $\lambda_{FM(PM)}$, and thickness $d_{FM(PM)}$, where the FM (PM) layer is in the range of $-d_{FM} \leq z \leq 0$ ($0 \leq z \leq d_{PM}$). The boundary conditions are given by $j_s^{FM}(-d_{FM}) = 0$, $j_m(-d_{FM}) = 0$, $j_s^{PM}(d_{PM}) + j_s^{SHE} = 0$, $\nabla\mu_c = 0$ at the system edges, and $j_s^{FM}(0) + j_m(0) = j_s^{PM}(0) + j_s^{SHE}$, where $j_s^{SHE}$ and $j_s^{FM(PM)}(z)$ denote the spin current induced by the SHE and the spin current in FM (PM) along the $z$ direction, respectively. We describe the spin-magnon interconversions at the PM/FM interface ($z = 0$) as

$$j_s^{FM}(0) = G_s[\mu_s^{FM}(0) - \mu_s^{PM}(0)]/e, \tag{5}$$

$$j_m(0) = G_m[\mu_m(0) - \mu_s^{PM}(0)]/e, \tag{6}$$

where $\mu_s^{FM(PM)}(z)$ denotes the spin accumulation in FM (PM) [25,28,59]. $G_s$ ($G_m$) represents the conductance for the interconversion between conduction electrons spins in PM and FM (between conduction electron spins in PM and magnons in FM). Subsequently, we obtain

$$j_s^{FM} = -\frac{G_s \Gamma_{FM}}{(G_s + \Gamma_{FM})} \frac{\sinh\left(\frac{[d_{FM}+z]}{\lambda_{FM}}\right)}{\sinh\left(\frac{d_{FM}}{\lambda_{FM}}\right)} \frac{\mu_s^{PM}(0)}{e}, \tag{7}$$

$$j_s^{PM} = -\frac{\cosh\left(\frac{z}{\lambda_{PM}}\right)}{\cosh\left(\frac{d_{PM}}{\lambda_{PM}}\right)} j_s^{SHE} + \Gamma_{PM} \frac{\sinh\left(\frac{[d_{PM}-z]}{\lambda_{PM}}\right)}{\sinh\left(\frac{d_{PM}}{\lambda_{PM}}\right)} \frac{\mu_s^{PM}(0)}{e}, \tag{8}$$

$$j_m = -\frac{G_m \Gamma_m}{(G_m + \Gamma_m)} \frac{\sinh\left(\frac{[d_{FM}+z]}{\lambda_m}\right)}{\sinh\left(\frac{d_{FM}}{\lambda_m}\right)} \frac{\mu_s^{PM}(0)}{e}, \tag{9}$$

where $\Gamma_{FM} = (1 - P_{FM}^2) \frac{\sigma_{FM} \tanh\left(\frac{d_{FM}}{\lambda_{FM}}\right)}{2\lambda_{FM}}$, $\Gamma_{PM} = \frac{\sigma_{PM} \tanh\left(\frac{d_{PM}}{\lambda_{PM}}\right)}{2\lambda_{PM}}$, $\Gamma_m = \frac{\sigma_m \tanh\left(\frac{d_{FM}}{\lambda_m}\right)}{\lambda_m}$, and

$$\mu_{\rm s}^{\rm PM}(0) = -\frac{ej_{\rm s}^{\rm SHE}\left(1 - {\rm sech}(\frac{d_{\rm PM}}{\lambda_{\rm PM}})\right)}{\left[\Gamma_{\rm PM} + \frac{G_{\rm s}\Gamma_{\rm FM}}{(G_{\rm s}+\Gamma_{\rm FM})} + \frac{G_{\rm m}\Gamma_{\rm m}}{(G_{\rm m}+\Gamma_{\rm m})}\right]}. \tag{10}$$

The resulting temperature modulation $\delta T$ can be obtained by the one-dimensional heat equation,

$$\kappa \partial_z^2 \delta T = -\partial_z\left(-\frac{\Pi_{\rm s}}{2}j_{\rm s} + \Pi_{\rm SPE}j_{\rm m}\right), \tag{11}$$

where $\kappa$ is the thermal conductivity. As our PM/FM system is connected to a heat bath, i.e., the substrate, at the bottom of the film and opened to the air at the surface of the film, we used $\delta T = 0$ at $z = -d_{\rm FM}$ and $\partial_z \delta T = 0$ at $z = d_{\rm PM}$, where we omit the black ink layer because there is negligibly small heat current as the heat radiation loss from the top surface is not effective compared with bulk thermal conduction. In this boundary condition, the temperature modulation induced by the SdPE is given by

$$\delta T_{\rm SdPE} = -\frac{\Pi_{\rm s}}{2\kappa_{\rm FM}}\frac{G_{\rm s}\Gamma_{\rm FM}}{(G_{\rm s}+\Gamma_{\rm FM})}\frac{j_{\rm s}^{\rm SHE}\lambda_{\rm FM}\tanh(\frac{d_{\rm FM}}{2\lambda_{\rm FM}})\left(1-{\rm sech}(\frac{d_{\rm PM}}{\lambda_{\rm PM}})\right)}{\left[\Gamma_{\rm PM} + \frac{G_{\rm s}\Gamma_{\rm FM}}{(G_{\rm s}+\Gamma_{\rm FM})} + \frac{G_{\rm m}\Gamma_{\rm m}}{(G_{\rm m}+\Gamma_{\rm m})}\right]} \tag{12}$$

and that induced by the SPE is given by

$$\delta T_{\rm SPE} = \frac{\Pi_{\rm SPE}}{\kappa_{\rm FM}}\frac{G_{\rm m}\Gamma_{\rm m}}{(G_{\rm m}+\Gamma_{\rm m})}\frac{j_{\rm s}^{\rm SHE}\lambda_{\rm m}\tanh(\frac{d_{\rm FM}}{2\lambda_{\rm FM}})\left(1-{\rm sech}(\frac{d_{\rm PM}}{\lambda_{\rm PM}})\right)}{\left[\Gamma_{\rm PM} + \frac{G_{\rm s}\Gamma_{\rm FM}}{(G_{\rm s}+\Gamma_{\rm FM})} + \frac{G_{\rm m}\Gamma_{\rm m}}{(G_{\rm m}+\Gamma_{\rm m})}\right]}. \tag{13}$$

Here, it is noteworthy that $\Gamma_{\rm FM}$ and $\Gamma_{\rm m}$ depends on $d_{\rm FM}$. The contribution from the interfacial thermal resistance can be included by

$$\delta T_{\rm SdPE}^{\rm int} = -\frac{\Pi_{\rm s}j_{\rm s}^{\rm FM}\big|_{z=0}}{2\kappa_{\rm int}}, \tag{14}$$

assuming continuity of the heat current at the PM/FM interface and the FM layer. The same manner can be applied to the SPE.

The above model calculations show that the SdPE and SPE have quite similar $d_{\rm FM}$ dependence. Although the difference between the SdPE and SPE comes from the transport properties of conduction electron spins and magnons, such as the length scale, conductivity, and conversion efficiencies at the interface, and the SdPE and SPE coefficients, it is difficult to estimate the SdPE and SPE parameters simultaneously by fitting; a number of parameters have to be assumed for quantitative discussions (see below). The experimental results in Fig. 5(b) show that the $d_{\rm FM}$ dependence of $\Delta A_{\rm odd}/j_{\rm c}^{\rm Pt}$ in the Pt/CoFeB films has a characteristic length of ~10 nm, which is similar to or rather longer than $\lambda_{\rm FM}$ for CoFeB, obtained in spin-valve experiments at low temperatures [60–62]. As the diffusion length of magnons can be longer than that of electron spins owing to the difference in the scattering mechanisms [25,28,57,60,63,64], the observed $\Delta A_{\rm odd}/j_{\rm c}^{\rm Pt}$ signals may contain the contribution from the magnon-driven SPE. Nevertheless, it is still difficult to separate the SdPE and SPE contributions quantitatively because of the presence of unknown transport parameters.

To obtain a clue for the separation, we estimated the spin-to-heat conversion coefficient from the magnitude of the observed $\Delta A_{\rm odd}/j_{\rm c}^{\rm Pt}$ signal. First of all, we have to note that, to compare with the model calculation with the LIT results, the amplitude $j_{\rm c}^{\rm Pt}$ of the square wave should be converted into the amplitude of the first harmonic sinusoidal wave: $(4/\pi)j_{\rm c}^{\rm Pt}$. If we attributed the signal for the Pt/CoFeB film solely to the SdPE, we obtained $\Pi_{\rm s}/\kappa_{\rm CoFeB}$ = -0.0051 ± 0.0008 VKmW$^{-1}$ and $\lambda_{\rm CoFeB}$ = 9.4 ± 4.6 nm from the fitting using the experimental values of $\sigma_{\rm CoFeB} = 6.0 \times 10^5$ $\Omega^{-1}$m$^{-1}$ [see the inset to Fig.5(b)], $\sigma_{\rm Pt} = 3.8 \times 10^6$ $\Omega^{-1}$m$^{-1}$, which is estimated based on the short-circuit model, and $d_{\rm Pt}$ = 10 nm and the reference values of $P_{\rm FM}$ = 0.72 [58], $\lambda_{\rm Pt}$ = 2 nm [48], and $\theta_{\rm SH}$=0.2 [65], where the fitting result is shown with a red solid line in Fig. 5(b). Here, $\kappa_{\rm CoFeB}$ and $\lambda_{\rm CoFeB}$ are the thermal conductivity and spin-diffusion length of CoFeB, respectively, and we assume an infinitely large $G_{\rm s}$, the condition in which $\mu_{\rm s}$ is continuous at the PM/FM interface, and the lower limit of $\Pi_{\rm s}$ is obtained. If $\kappa_{\rm CoFeB}$ is comparable to the thermal conductivity of CoFe, i.e., assuming $\kappa_{\rm CoFeB}$ = 29.8 Wm$^{-1}$K$^{-1}$ [44], we obtained $\Pi_{\rm s}$ = -0.152 ± 0.023 V. The $\Pi_{\rm s}/\kappa_{\rm Py}$ value for Py is estimated to be -0.00052 ± 0.00020 VKmW$^{-1}$ from the experimental values of $\sigma_{\rm Py} = 1.3 \times 10^6$ $\Omega^{-1}$m$^{-1}$ and $d_{\rm Py}$ = 20 nm, the reference values of $P_{\rm FM}$ = 0.36 [44] and $\lambda_{\rm Py}$ = 6.7 nm [57], and the aforementioned parameters for Pt, indicating $\Pi_{\rm s}$ = -0.0119 ± 0.0045 V when the thermal conductivity of Py is $\kappa_{\rm Py}$ = 22.9 Wm$^{-1}$K$^{-1}$ [66]. We note that the injection efficiency of the conduction-electron spin current, $j_{\rm s}^{\rm FM}(0)/j_{\rm s}^{\rm SHE}$, for the Pt/CoFeB(20 nm) [Pt/Py(20 nm)] interface is as low as 1.5 % [8.0 %] because of the huge difference between $\sigma_{\rm CoFeB}$ and $\sigma_{\rm Pt}$. Therefore, although the magnitude of the observed temperature modulation in the Pt/CoFeB systems is comparable to that in the Pt/Py systems, the estimated $\Pi_{\rm s}$ value for the Pt/CoFeB systems is much greater than that for the Pt/Py systems. Notably, the estimated $\Pi_{\rm s}$ values are much greater than the reported values of the SdPE coefficients, -0.0216 V for CoFeAl [44], -0.0059 V for CoFe [42], and -0.0011 V [42] and -0.0019 V [31,35,37] for Py, and even greater than the conventional (spin-independent) Peltier coefficients for CoFeB and Py [7,44], where the SdPE coefficients are estimated by multiplying the SdSE coefficients by the temperature through the Onsager reciprocal relation [5,37]. This situation remains even when taking the contribution from the interfacial thermal resistance of the PM/FM junctions into account; assuming $\kappa_{\rm int}$ = 1 GWm$^{-2}$K$^{-1}$ as a typical value of the interfacial thermal conductance for metal-metal junctions [67–69], we obtained $\Pi_{\rm s}$ = -0.062 ± 0.022 V and $\lambda_{\rm CoFeB}$ = 18.1 ± 9.9 nm for the Pt/CoFeB systems, where the fitting result with $\kappa_{\rm int}$ is shown with a red dotted line in Fig. 5(b). These facts indicate that the results cannot be explained only by the electron-driven SdPE due to the SHE, indicating the substantial contribution from the magnon-driven SPE even in the metallic systems. In fact, the magnitude of the spin-current-induced temperature modulation, $\Delta A_{\text{odd}}/j_c^{\text{Pt}}$, in Pt/CoFeB ($0.07 \times 10^{-12}$ Km$^2$A$^{-1}$ for $d_{\text{CoFeB}} = 20$ nm) and Pt/Py ($0.03 \times 10^{-12}$ Km$^2$A$^{-1}$ for $d_{\text{Py}} = 20$ nm) films is comparable to that of the SPE in the Pt/Fe$_3$O$_4$ system ($0.13 \times 10^{-12}$ Km$^2$A$^{-1}$ for the 23-nm-thick Fe$_3$O$_4$ layer) [39].

Finally, we mention remaining tasks for realizing quantitative estimation of the spin-to-heat conversion phenomena in metallic systems. As discussed above, the temperature modulation induced by the SdPE and SPE is determined by many transport parameters in PM/FM systems, and it is necessary to determine their reliable values with the aid of other experiments and calculations. Furthermore, in the PM/FM bilayer systems, thermo-spin and/or thermoelectric conversion due to the interfacial effects may have to be taken into account. For example, the spin current due to the spin anomalous Hall effect in the FM layers [70] can generate the SdPE signal and its output can be modified when the spin-sink PM layer is attached. This contribution is hard to be separated from other effects but is expected to be small because the modulation of the spin anomalous Hall effect cannot explain the sign change of the spin-current-induced temperature modulation between the Pt/CoFeB and W/CoFeB systems. Another possibility is the enhancement of the AEE due to the interfacial spin-orbit interaction, because the anomalous Nernst effect, the reciprocal of the AEE, was observed to be enhanced in PM/FM multilayer films with increasing the PM/FM-interface density [71]. However, such interfacial effect can be ruled out by the $d_{\text{FM}}$ dependence of the temperature modulation since the interfacial contribution is expected to decrease with increasing $d_{\text{FM}}$, which is an opposite trend to the results shown in Fig. 5(b).

## IV. CONCLUSION

In this paper, we reported the measurements of the temperature modulation induced by thermoelectric and thermo-spin effects in PM(Pt, W, or Ta)/FM(CoFeB or Py) bilayer films and FM monolayer films by means of the lock-in thermography technique. We observed clear temperature-modulation signals satisfying the symmetry of the SPE, SdPE, and AEE and found that all the PM/FM bilayer films exhibit finite spin-current-induced contributions, which are estimated by subtracting the AEE contribution in FM. The sign and the PM-thickness dependence of the spin-current-induced temperature modulation are consistent with the interpretation that the temperature modulation is driven by the SHE in PM. The CoFeB-thickness dependence of the spin-current-induced temperature modulation in the Pt/CoFeB films suggests that the length scale of the observed phenomenon is in the order of 10 nm. Importantly, the magnitude of the spin-current-induced temperature modulation in our PM/FM bilayer films is too large to be explained only by the SdPE contribution, indicating that both the electron-driven SdPE and the magnon-driven SPE contribute to the temperature modulation in our films. This fact is revealed owing to the versatility of the LIT method, which allows us to overcome the difficulty in conventional temperature measurements in micro-fabricated nanoscale devices. Although the quantitative separation between the SPE and SdPE contributions remains to be achieved, the observation of the spin-to-heat current conversion in simple metallic bilayers makes significant progresses in the physics of spin caloritronics.

## ACKNOWLEDGMENTS

The authors thank R. Tsuboi for his assistance in the analyses of the LIT images and S. Isogami, K. Yamanoi, and T. Kimura for valuable discussions. This work was supported by PRESTO "Phase Interfaces for Highly Efficient Energy Utilization" (JPMJPR12C1), CREST "Creation of Innovative Core Technologies for Nano-enabled Thermal Management" (JPMJCR17I1), and ERATO "Spin Quantum Rectification Project" (JPMJER1402) from JST, Japan; Grant-in-Aid for Scientific Research (A) (JP15H02012), Grant-in-Aid for Scientific Research on Innovative Area "Nano Spin Conversion Science" (JP26103005), and Grant-in-Aid for Specially Promoted Research (JP15H05702) from JSPS KAKENHI, Japan; the Inter-University Cooperative Research Program of the Institute for Materials Research, Tohoku University (17K0005); and the NEC Corporation. S.D. is supported by JSPS through a research fellowship for young scientists (JP16J02422).

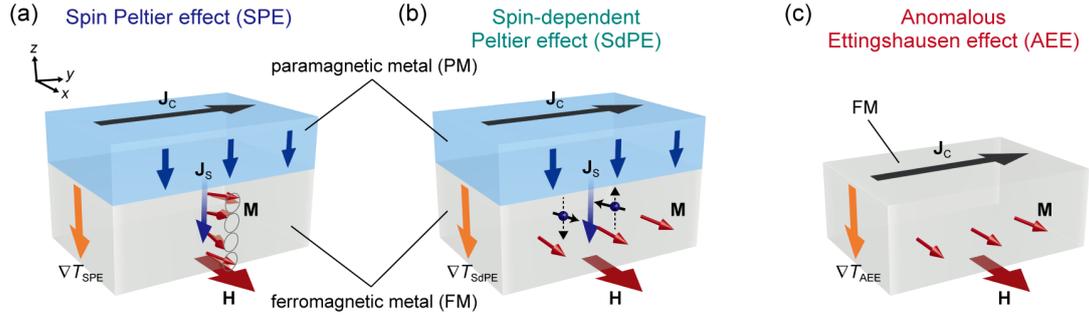

FIG. 1 Schematic illustrations of (a) the spin Peltier effect (SPE) driven by the spin Hall effect (SHE), (b) the spin-dependent Peltier effect (SdPE) driven by the SHE, and (c) the anomalous Ettingshausen effect (AEE). **H**, **M**, **J**$_c$, and **J**$_s$ denote the magnetic field vector with the magnitude $H$, magnetization vector with the magnitude $M$ of a ferromagnetic metal (FM), charge current, and spatial direction of the spin current generated by the SHE in a paramagnetic metal (PM), respectively. $\nabla T_{SPE}$, $\nabla T_{SdPE}$, and $\nabla T_{AEE}$ represent the temperature gradient appearing as a result of the heat current induced by the SPE, SdPE, and AEE, respectively.

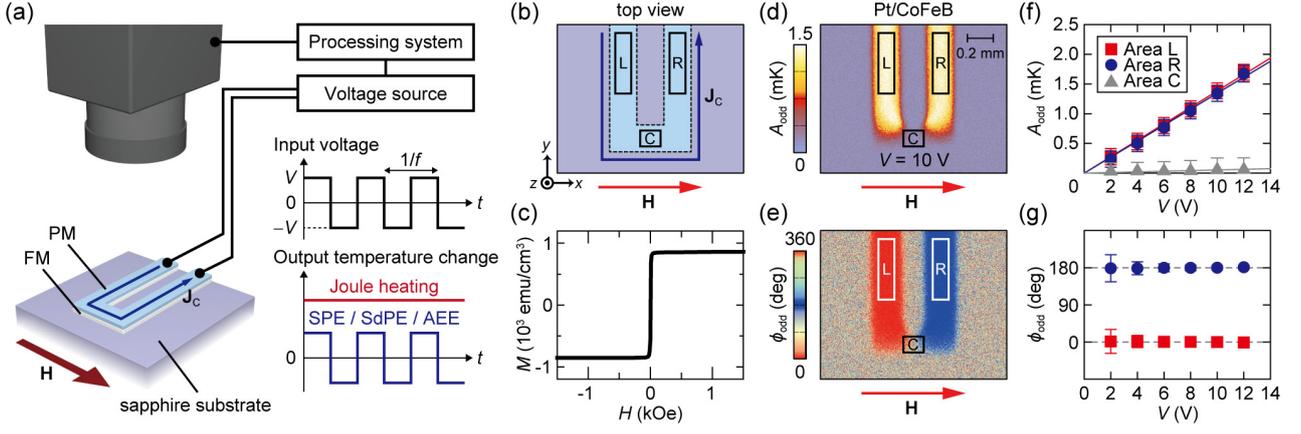

FIG. 2 (a) Lock-in thermography (LIT) for the measurements of the SPE, SdPE, and AEE in the PM/FM bilayer systems. $V$ and $f$ denote the amplitude and frequency of the rectangularly-modulated AC voltage applied to the PM/FM film. (b) Schematic of the sample system from the top view. The squares on the PM/FM film define the areas L, R, and C. (c) $M$-$H$ curve for a 20-nm-thick CoFeB film on a sapphire substrate, where the $H$-linear contribution from the substrate was subtracted from raw data. (d),(e) $A_{\text{odd}}$ and $\phi_{\text{odd}}$ images for the Pt/CoFeB film at $V = 10$ V and $|H| = 1.4$ kOe, where $A_{\text{odd}}$ ($\phi_{\text{odd}}$) denotes the lock-in amplitude (phase) of the temperature modulation with the $H$-odd dependence. The thickness of the Pt (CoFeB) layer is 10 nm (20 nm). (f) $V$ dependence of $A_{\text{odd}}$ on L, R, and C of the Pt/CoFeB film, where the plotted data were obtained by averaging the $A_{\text{odd}}$ values on the areas. (g) $V$ dependence of $\phi_{\text{odd}}$ on L and R of the Pt/CoFeB film.

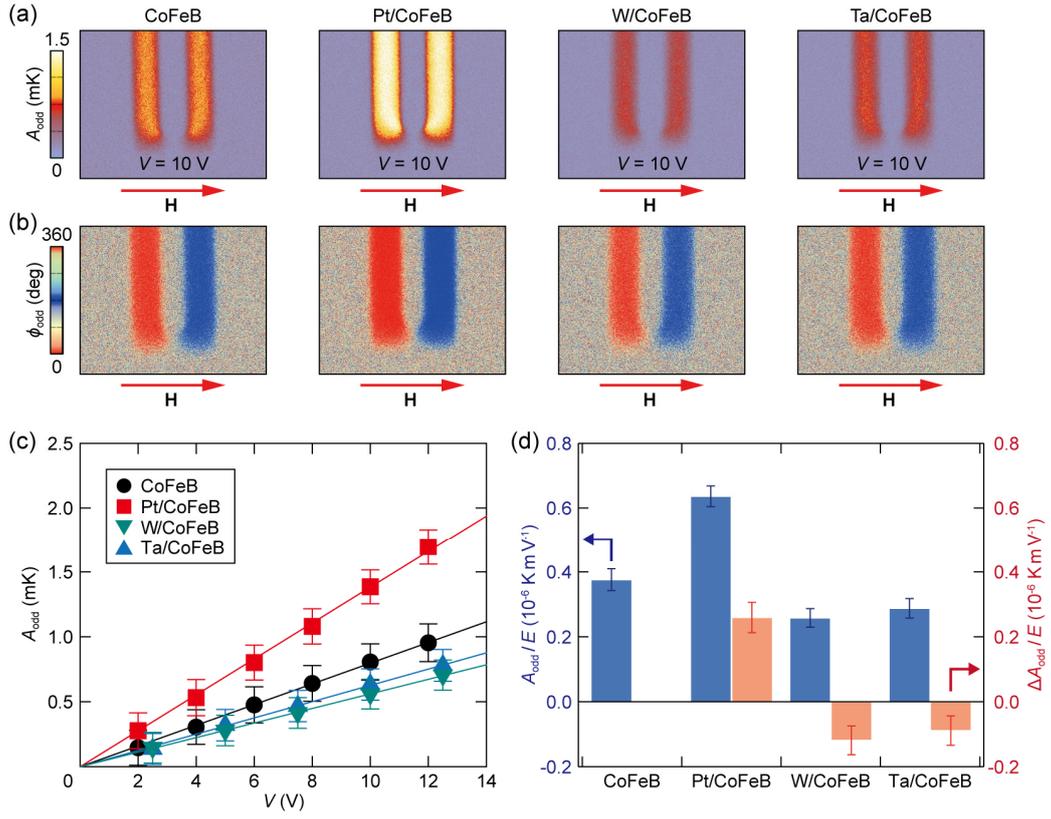

FIG. 3 (a),(b) $A_{odd}$ and $\phi_{odd}$ images for the CoFeB monolayer and PM(Pt, W, or Ta)/CoFeB bilayer films at $V$ = 10 V. (c) $V$ dependence of $A_{odd}$ on the area L of the CoFeB and PM/CoFeB films. (d) $A_{odd}/E$ and $\Delta A_{odd}/E$ values on L of the CoFeB and PM/CoFeB films. The $\Delta A_{odd}$ value was obtained by subtracting the $A_{odd}$ value averaged over L of the CoFeB film from that of the PM/CoFeB film.

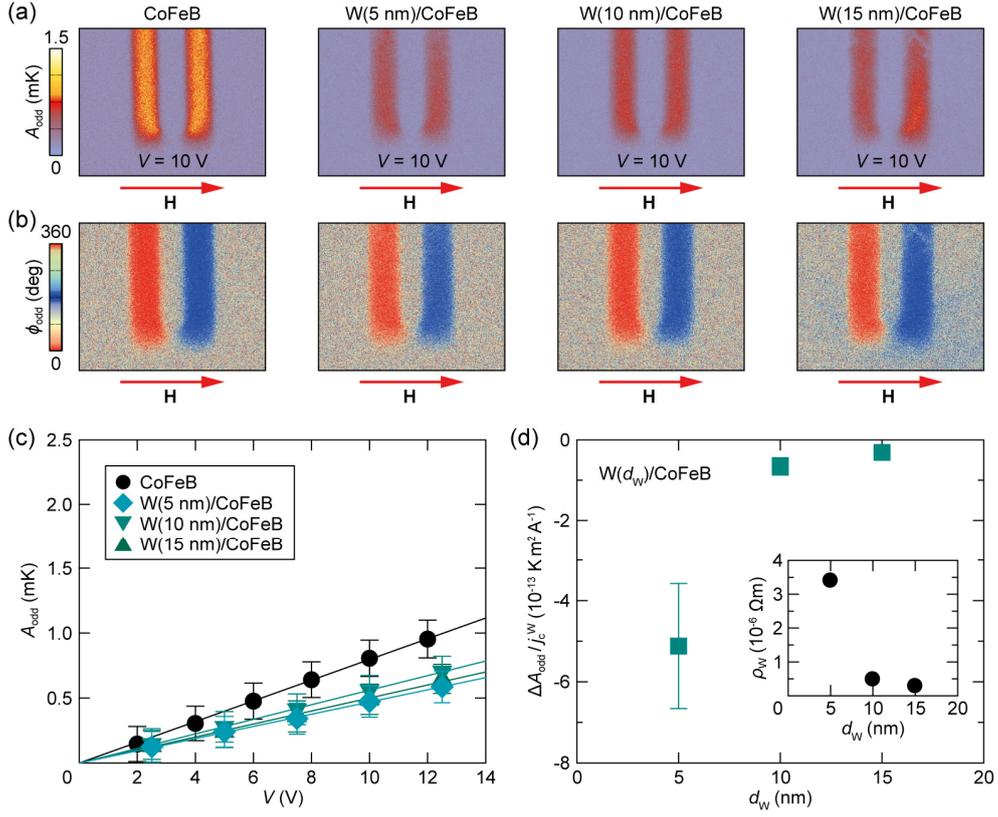

FIG. 4 (a),(b) $A_{odd}$ and $\phi_{odd}$ images for the CoFeB monolayer and W($d_W$)/CoFeB bilayer films with different W-layer thicknesses, $d_W = $ 5, 10, and 15 nm, at $V = 10$ V and $|H| = 0.7$ kOe. (c) $V$ dependence of $A_{odd}$ on the area L of the CoFeB and W($d_W$)/CoFeB films. (d) $d_W$ dependence of $\Delta A_{odd}/j_c^W$ on L of the W($d_W$)/CoFeB films, where the charge-current density $j_c^W$ in the W layer was estimated based on the parallel circuit model [53]. The inset to (d) shows the $d_W$ dependence of the electrical resistivity $\rho_W$ of the W layer.

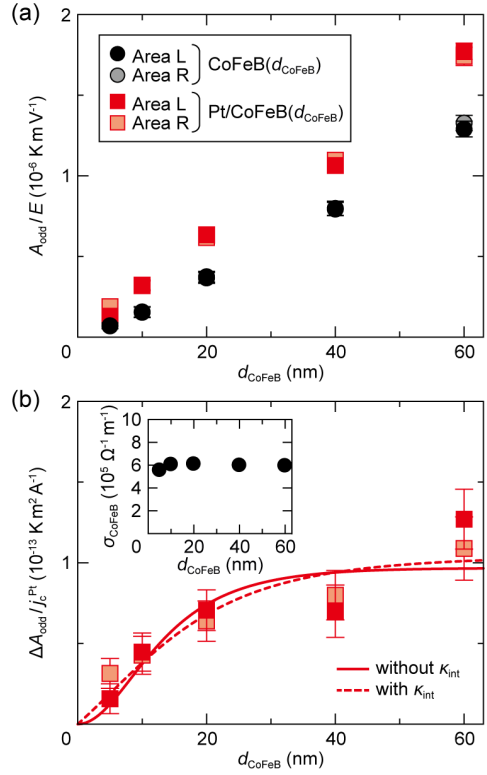

FIG. 5 (a) CoFeB-thickness $d_{CoFeB}$ dependence of $A_{odd}/E$ on the areas L and R of the CoFeB monolayer and Pt/CoFeB bilayer films. The $A_{odd}/E$ values are estimated by linear fitting of the $V$ dependence of $A_{odd}$. (b) $d_{CoFeB}$ dependence of $\Delta A_{odd}/j_c^{Pt}$ on L and R. The $\Delta A_{odd}$ value was obtained by subtracting the $A_{odd}$ value averaged over L or R of the CoFeB($d_{CoFeB}$) film from that of the Pt/CoFeB($d_{CoFeB}$) films. The solid (dashed) fitting curve is obtained by fitting the experimental results using Eq. (12) [Eqs. (12) and (14)] for the case without (with) the interfacial thermal conductance. Parameters used in the fitting are $\kappa_{CoFeB} = 29.8$ Wm$^{-1}$K$^{-1}$ [44], $\sigma_{CoFeB} = 6.0 \times 10^5$ $\Omega^{-1}$m$^{-1}$, $\sigma_{Pt} = 3.8 \times 10^6$ $\Omega^{-1}$m$^{-1}$, $d_{CoFeB} = 20$ nm, $d_{Pt} = 10$ nm, $P_{FM}=0.72$ [58], $\lambda_{Pt} = 2$ nm [48], and $\theta_{SH}=0.2$ [65]. Here, we use $G_m = 0$ $\Omega^{-1}$m$^{-2}$ to exclusively consider the SdPE contribution and an infinitely large $G_s$ value to assume the condition that $\mu_s$ is continuous at the PM/FM interface. For the case without the interfacial thermal conductance, $\Pi_s = -0.152 \pm 0.023$ V and $\lambda_{CoFeB} = 9.4 \pm 4.6$ nm are obtained. For the other case with $\kappa_{int} = 1$ GWm$^{-2}$K$^{-1}$, $\Pi_s = -0.062 \pm 0.022$ V and $\lambda_{CoFeB} = 18.1 \pm 9.9$ nm are obtained. The inset to (a) shows the $d_{CoFeB}$ dependence of the electrical conductivity $\sigma_{CoFeB}$ of the CoFeB films.

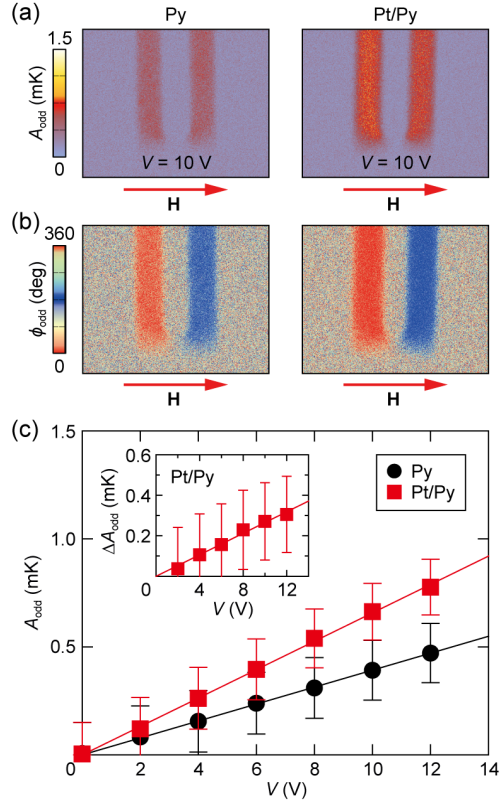

FIG. 6 (a),(b) $A_{\text{odd}}$ and $\phi_{\text{odd}}$ images for the Py monolayer and Pt/Py bilayer films at $V = 10$ V and $|H| = 1.4$ kOe. (c) $V$ dependence of $A_{\text{odd}}$ on L of the Py and Pt/Py films. The inset to (c) shows the $V$ dependence of $\Delta A_{\text{odd}}$, where the $\Delta A_{\text{odd}}$ value was obtained by subtracting the $A_{\text{odd}}$ value averaged over L of the Py film from that of the Pt/Py film.